\UseRawInputEncoding

\documentclass[pdflatex,sn-chicago]{sn-jnl}

\jyear{2023}%

\theoremstyle{thmstyleone}%
\theoremstyle{thmstyletwo}%

\theoremstyle{thmstylethree}%

\newcommand{\argmin}{\mathop{\rm arg~min}\limits}
\providecommand{\abs}[1]{\lvert#1\rvert}

\usepackage{bm}
\usepackage{etoolbox} 

\makeatletter
\patchcmd{\ps@headings}
{\hbox to \hsize{\hfill Springer Nature 2021 \LaTeX\ template\hfill}}
{\hbox to \hsize{\hfill \hfill}}
{}
{}
\makeatother
\makeatletter
\patchcmd{\ps@headings}
{\hbox to \hsize{\hfill Springer Nature 2021 \LaTeX\ template\hfill}}
{\hbox to \hsize{\hfill \hfill}}
{}
{}
\makeatother
\makeatletter
\patchcmd{\ps@headings}
{\hbox to \hsize{\hfill Springer Nature 2021 \LaTeX\ template\hfill}}
{\hbox to \hsize{\hfill \hfill}}
{}
{}
\makeatother

\begin{document}

\title[]{Multi-Task Learning Regression via Convex Clustering}


\author*[1]{\fnm{Akira} \sur{Okazaki}}\email{okazaki.akira.864@s.kyushu-u.ac.jp}

\author[2]{\fnm{Shuichi} \sur{Kawano}}

\affil[1]{\orgdiv{Graduate School of Mathematics}, \orgname{Kyushu University}, \orgaddress{\street{744 Motooka}, \city{Nishi-ku}, \state{Fukuoka}, \postcode{819-0395} \country{Japan}}}
\affil[2]{\orgdiv{Faculty of Mathematics}, \orgname{Kyushu University}, \orgaddress{\street{744 Motooka}, \city{Nishi-ku}, \state{Fukuoka}, \postcode{819-0395} \country{Japan}}}


\abstract{
Multi-task learning (MTL) is a methodology that aims to improve the general performance of estimation and prediction by sharing common information among related tasks. In the MTL, there are several assumptions for the relationships and methods to incorporate them. One of the natural assumptions in the practical situation is that tasks are classified into some clusters with their characteristics. For this assumption, the group fused regularization approach performs clustering of the tasks by shrinking the difference among tasks. This enables us to transfer common information within the same cluster. However, this approach also transfers the information between different clusters, which worsens the estimation and prediction.
To overcome this problem, we propose an MTL method with a centroid parameter representing a cluster center of the task. Because this model separates parameters into the parameters for regression and the parameters for clustering, we can improve estimation and prediction accuracy for regression coefficient vectors.
We show the effectiveness of the proposed method through Monte Carlo simulations and applications to real data. 
}




\maketitle

\section{Introduction}\label{sec1}

Multi-task learning (MTL) \citep{Caruana1997-bb} is a statistical methodology that simultaneously estimates multiple models for each task. It aims to improve general estimation and prediction accuracy by transferring related information among tasks. If the tasks are sufficiently related, MTL can lead to better performance than independently estimating each task. Due to this advantage, MTL has been applied to many problems in various fields of research such as disease progression prediction \citep{Zhou2011-wm}, biomedicine \citep{Li2018-tf}, transportation \citep{Deng2017-oc}, image annotation \citep{Fan2008-ad}, speech recognition \citep{Parameswaran2010-of}, and so on.

MTL methods are roughly classified into two approaches according to the assumption of relationships among tasks. The first is to assume that all tasks share a common structure. This approach is achieved by estimating low-rank representation \citep{Ando2005-pl}, sparsity pattern \citep{Obozinski2010-rt}, and so on. However, in some practical situations, it is difficult to assume that all tasks have the same structure. If there are tasks with different characteristics, this approach fails to transfer common information. Meanwhile, the second approach is to assume that tasks with similar characteristics form multiple clusters and to aim to learn underlying task groups. This approach is achieved by clustering the task's parameters in order to transfer the information characterized by each cluster \citep{Kang2011-gs}. For this approach, MTL methods with the group fused $\ell_{q}$-norm regularization have been proposed (\cite{Yamada2017-cg}; \cite{He2019-wy}; \cite{Dondelinger2020-wz}; \cite{Zhang2022-lt}). These methods perform clustering by shrinking the difference between parameters for each task. Because they are formulated as convex optimization problems, the global optima of the parameters can be obtained. However, the regularization term shrinks the difference of parameters for irrelevant tasks that should belong to different clusters. This means that the incorrect transfer of information between different characteristics is caused. As a result, this worsens the estimation and prediction accuracy.

To overcome this problem, we propose an MTL regression method with centroid parameters representing the cluster center of each task. 
Instead of shrinking the differences of regression coefficient vectors, those of centroid parameters are shrunk. In addition, the value of the regression coefficient vector is estimated around the value of the corresponding centroid parameter. Thus, the shrinkages between the regression coefficient vectors for irrelevant tasks are expected to be reduced. We employ squared $\ell_{2}$-norm to regularize the regression coefficient vectors and $\ell_{2}$-norm to cluster the centroids, which are based on convex clustering (\cite{Pelckmans2005-kw}; \cite{Hocking2011-mn}; \cite{Lindsten2011-bs}). This leads to keeping the proposed method as a convex optimization problem. The parameters are estimated by the block coordinate descent algorithm, which is performed by alternately optimizing regularized regression and convex clustering. 

This paper is organized as follows. In Section \ref{Sec2}, we review the MTL method based on group fused $\ell_{q}$-norm and the convex clustering. In Section \ref{Sec3}, we propose MTLCVX and describe existing work. In Section \ref{Sec4}, we provide an estimation algorithm for MTLCVX. Monte Carlo simulations and application to real data are illustrated in Sections \ref{Sec5} and \ref{Sec6}. The concluding remarks are given in Section \ref{Sec7}.

\section{Method}
\label{Sec2}
\subsection{Multi-task learning based on group fused regularization}

Suppose that we have $n_{m}$ observed $p$-dimensional data $\{\bm{x}_{mi};i=1,\ldots,n_{m}\}$ for the explanatory variables and $n_{m}$ observed data $\left\{y_{mi};i=1,\ldots,n_{m}\right\}$ for the response variable from the $m$-th task $(m=1,\ldots,T)$.
These pairs $\{({y}_{mi},\bm{x}_{mi});i=1,\ldots,n_{m}\}$ are given independently. We set $\bm{y}_{m}=(y_{m1},\ldots,y_{mn_{m}})^{\top}\in\mathbb{R}^{n_{m}}$ and 
$X_{m}=(\bm{x}_{m1},\ldots \bm{x}_{mn_{m}})^{\top}\in\mathbb{R}^{n_{m}\times p}$, where $\bm{y}_{m}$ is assumed to be centered with zero mean, and each $\bm{x}_{mi}$ is assumed to be standardized with zero mean and unit variance. Furthermore, we assume that $\left\{ (\bm{y}_{m},X_{m}); m=1,\ldots,T\right\}$ are given for $T$ tasks. 

For these $T$ tasks, we consider the following multiple regression models:
    \begin{equation}
    \label{linearMTL}
    \bm{y}_{m} = \bm{X}_{m}\bm{w}_{m}+ \bm{\epsilon}_{m} , \quad \quad m=1,\ldots,T,
    \end{equation}
    where $\bm{w}_{m}=(w_{m1},\ldots,w_{mp})^{\top}$ is a regression coefficient vector for $m$-th task and $\bm{\epsilon}_{m}$ is an error term whose elements distributed as $N(0,\sigma^{2})$ independently. Note that intercepts are excluded from the model since we assume the response vector is centered and explanatory variables are standardized for each task.
    For Model (\ref{linearMTL}), we consider the following minimization problem:
    \begin{equation}
        \label{groupregMTL}
        \min_{\substack{\bm{w}_{m}\in \mathbb{R}^{p}\\m=1,\ldots,T}}
        \left\{ \sum_{m=1}^{T} \frac{1}{2n_{m}}\|\bm{y}_{m} - \bm{X}_{m}\bm{w}_{m} \|_{2}^{2}+\lambda\sum_{(m,l)\in\mathcal{E}}r_{m,l}\|\bm{w}_{m}-\bm{w}_{l} \|_{q} \right\},
    \end{equation}
    where $r_{m,l}$ is a weight between $m$-th and $l$-th task, $\mathcal{E}$ is a set of task pairs $(m,l)$, and $\lambda\;(\geq 0)$ is a regularization parameter. The first term is a loss function of the linear regression model and the second term is a group fused $\ell_{q}$-norm regularization term. This second term induces similarity among tasks by estimating $\bm{w}_{m} \simeq \bm{w}_{l}$. When $\lambda=0$, Problem (\ref{groupregMTL}) is the same as independently estimating ordinal linear regression models. However, when $\lambda>0$, this minimization problem performs MTL by sharing common information among other related tasks. When $q\geq1$, Problem (\ref{groupregMTL}) becomes a convex optimization problem, and then a global minimum can be obtained. When $q=2$, this model is in a framework of a network lasso \citep{Hallac2015-ss}. We can solve the model by its estimation algorithm based on the alternating direction method of multipliers (ADMM) \citep{Boyd2011-ul}.
    
    \cite{Yamada2017-cg} considered the situation for $q=2$ and $n_{m}=1$. They performed variable selection by adding $\ell_{1,2}$-norm regularization. On the other hand, to deal with situations where the number of tasks is more than ten thousand, \cite{He2019-wy} proposed the efficient and scalable estimation algorithm for Problem (\ref{groupregMTL}).
    
    \subsection{Convex clustering}
    
    Suppose that we have $n$ observed $p$-dimensional data $\left\{\bm{x}_{i};i=1,\ldots,n\right\}$. To classify these data into $C$ exclusive clusters, convex clustering (\cite{Pelckmans2005-kw}; \cite{Hocking2011-mn};  \cite{Lindsten2011-bs}) has been proposed. Convex clustering is formulated by the following minimization problem:
     \begin{equation}
        \label{CVXC}
        \min_{\bm{u}_{i}\in \mathbb{R}^{p},\:i=1,\ldots,n}
        \left\{ \sum_{i=1}^{n} \frac{1}{2}\|\bm{x}_{i} - \bm{u}_{i} \|_{2}^{2}+\lambda_{1}\sum_{(i,j)\in\mathcal{E}}r_{i,j}\|\bm{u}_{i}-\bm{u}_{j} \|_{q} \right\},\\
    \end{equation}
  where $\bm{u}_{i}\in\mathbb{R}^{p}$ is a parameter vector for $i$-th sample, which is called a centroid. If the value of $\bm{u}_{i}$ and $\bm{u}_{j}$ are estimated to be the same by the second term, corresponding samples $\bm{x}_{i}$ and $\bm{x}_{j}$ are considered as belonging to the same cluster. To shrink the difference between $\bm{u}_{i}$ and $\bm{u}_{j}$ into exactly zero, $q=1,2$ and $\infty$ are often used.
  Because the convex clustering is also viewed as a convex relaxation of a $k$-means \citep{Tan2015-oo},
 a centroid $\bm{u}_{i}$ is considered as a biased cluster center in the $k$-means, which means that the estimated centroids $\widehat{\bm{u}}_{i}$ are affected by shrinkage with other cluster's centroids.
 The value of weights $r_{i,j}$ is calculated by a $k$-nearest neighbor and a Gaussian kernel empirically (\cite{Lindsten2011-bs}; \cite{Sun2021-jt}). On the other hand, each iteration in the updates of Problem (\ref{CVXC}) contains a computation for the second term for all combinations of samples. Therefore, the computational cost in each iteration is $\mathcal{O}(n^2)$. However, by using $k$-nearest neighbor, many weights are set to zero. Then, the cost is reduced to $\mathcal{O}(kn)$ \citep{Sun2021-jt}.
\section{Proposed method}
\label{Sec3}
\subsection{Multi-task learning via convex clustering}

The second term of Model (\ref{groupregMTL}) has the problem that a task is affected by other tasks belonging to other clusters as with the convex clustering. To reduce the problem, \cite{Yamada2017-cg} and \cite{He2019-wy} calculated the weights $r_{m,l}$ using $k$-nearest neighbor. \cite{Zhou2016-ik} and \cite{Shimamura2021-ma} proposed the methods that treat the weights as latent parameters, and estimated these parameters and regression coefficient parameters simultaneously. Because the latter approach induces the non-convexity of the model, it is difficult to construct the estimation algorithm converging into the global minimum.

 To overcome this problem, we propose the following problem:

\begin{equation}
    \begin{split}
        \label{prop1} \min_{\substack{\bm{w}_{m},\bm{u}_{m}\in \mathbb{R}^{p}\\m=1,\ldots,T}}
                \left\{ \sum_{m=1}^{T} L(\bm{w}_{m},w_{m0})+\frac{\lambda_{1}}{2}\sum_{m=1}^{T}\|\bm{w}_{m}-\bm{u}_{m}\|_{2}^{2}+\lambda_{2}\sum_{(m,l)\in\mathcal{E}}r_{m,l}\|\bm{u}_{m}-\bm{u}_{l} \|_{2} \right\}, \\
        \end{split}
\end{equation}
where $\bm{u}_{m}\in\mathbb{R}^{p}$ is a centroid for $m$-th task, $\lambda_{1}$ and $\lambda_{2}$ are tuning parameters with non-negative value, $L(\bm{w}_{m},w_{m0})$ is a loss function for $m$-th task. The second term is a squared-$\ell_{2}$ norm to estimate  
the value of $\bm{w}_{m}$ around $\bm{u}_{m}$. The third term is a $\ell_{2}$-norm in order to perform the clustering of $\bm{u}_{m}$. For the loss function, when continuous response vectors $\bm{y}_{m}\in\mathbb{R}^{n_{m}}$ are considered, we adopt the squared loss function:
\begin{equation}
\label{loss_linear}
    L(\bm{w}_{m},w_{m0}) = \frac{1}{2n_{m}}\|\bm{y}_{m} - \bm{X}_{m}\bm{w}_{m} \|_{2}^{2}.
\end{equation}
Meanwhile, when binary response vectors $\bm{y}_{m}\in\{0,1\}^{n_{m}}$ are considered, we adopt the logistic loss function
\begin{equation}
\label{loss_losistic}
    L(\bm{w}_{m},w_{m0}) = - \frac{1}{n_
{m}}\sum_{i=1}^{n_{m}}\left\{y_{mi}(w_{m0}+\bm{w}_{m}^{\top}\bm{x}_{i})-\log\{1+{\exp{(w_{m0}+\bm{w}_{m}^{\top}\bm{x}_{i})}}\} \right\},
\end{equation}
where $w_{m0}$ is a intercept for $m$-th task. Note that only when $L(\bm{w}_{m},w_{m0})$ is a loss function of the linear regression, the intercepts are excluded from the model without a loss of generality. 

In Problem \eqref{prop1}, the regression coefficient vectors $\bm{w}_{m}$ are not shrunk directly unlike Problem (\ref{groupregMTL}), while $\bm{u}_{m}$ are shrank and clustered. When the value of $\lambda_{1}$ is large, $\bm{w}_{m}$ is estimated to be the same value of $\bm{u}_{m}$, which is close to Problem (\ref{groupregMTL}). However, when the value of $\lambda_{1}$ is small, the value of $\bm{w}_{m}$ can be estimated to be different from that of $\bm{u}_{m}$. Therefore, we can expect to reduce the shrinkage among irrelevant tasks. 

The proposed method is a jointly convex optimization problem with respect to $\bm{w}_{m}$ and $\bm{u}_{m}$. This is readily confirmed by the following two calculations. The first is that the Hessian matrix of the sum of the first term and the second term is a semi-positive definite for both the loss function of linear regression (\ref{loss_linear}) and logistic regression (\ref{loss_losistic}). Next, the third term is a convex function in general. Thus, the sum of these terms is a convex optimization problem.

Because the second and third terms are viewed as regularization terms derived from the model of convex clustering, 
we refer to this model as MTLCVX (\textbf{M}ulti-\textbf{T}ask \textbf{L}earning via \textbf{C}on\textbf{V}e\textbf{X} clustering).

We set the weights $r_{m,l}$ in \eqref{prop1} as in \cite{Yamada2017-cg}:
    \begin{equation}
        \label{setR}
            R = \frac{S^{\top}+S}{2}, \qquad(S)_{ml}=
            \begin{cases}
                 & 1 \quad \widehat{\bm{w}}^{\mathrm{SL}}_{m}\;\mathrm{is}\; \mathrm{a}\;k\mathrm{-nearest}\; \mathrm{neighbor} \; \mathrm{of}\;\widehat{\bm{w}}_{l}^{\mathrm{SL}}, \\
                 & 0 \quad \mathrm{otherwise},
            \end{cases}
    \end{equation} 
    where $\widehat{\bm{w}}^{\mathrm{SL}}_{m}$ is an estimated regression coefficient vector for $m$-th task by single-task learning such as the OLS, ridge, and lasso.
    From this equation, if $r$-th task and $m$-th task are $k$-nearest neighbors of each other, then $r_{m,l}=1$. If they are $k$-nearest neighbors from only one side, then $r_{m,l}=0.5$. While \cite{He2019-wy} only set $r_{m,l}=\{0,1\}$ in a similar way, Eq. (\ref{setR}) may allow us to reduce the effects of false-positive weights.

    \subsection{Multi-task learning via adaptive convex clustering }
    A drawback of Eq. (\ref{setR}) is that weights $r_{m,l}$ may have some noises, since the estimated value $\widehat{\bm{w}}^{\mathrm{SL}}_{m}$ may not be accurate.
   To address it, we consider calculating weights $r_{m,l}$ as in the adaptive lasso \citep{Zou2006-yg}: 
    \begin{equation}
        \label{prop2}
            \min_{\substack{\bm{w}_{m},\bm{u}_{m}\in \mathbb{R}^{p},\\ \:m=1,\ldots,T}}
            \left\{ \sum_{m=1}^{T} L(\bm{w}_{m},w_{m0})+\frac{\lambda_{1}}{2}\sum_{m=1}^{T}\|\bm{w}_{m}-\bm{u}_{m}\|_{2}^{2}+\lambda_{2}\sum_{(m,l)\in\mathcal{E}}\widehat{r}_{m,l}\|\bm{u}_{m}-\bm{u}_{l} \|_{2} \right\},
\end{equation}
where $\widehat{r}_{m,l}$ is an adaptive weight. This weight is computed as follows:
\begin{equation}
    \label{scale}
    \begin{split}
    \widehat{r}_{m,l}&=\frac{1}{\|\widehat{\bm{u}}_{m}(\mathrm{MTLCVX})-\widehat{\bm{u}}_{l}(\mathrm{MTLCVX}) \|_{2}}\nu,\\
         \nu&=\left(\sum_{(m,l)\in\mathcal{E}}\frac{1}{\|\widehat{\bm{u}}_{m}(\mathrm{MTLCVX})-\widehat{\bm{u}}_{l}(\mathrm{MTLCVX})\|_{2}}\right)^{-1}\sum_{(m,l)\in\mathcal{E}}r_{m,l},
    \end{split}
\end{equation}
where $\widehat{\bm{u}}_{m}(\mathrm{MTLCVX})$ is an estimated value of a centroid $\bm{u}_{m}$ in Problem (\ref{prop1}), and $\nu$ is a scaling parameter.
The scaling parameter $\nu$ is defined to ensure $\sum_{(m,l)\in\mathcal{E}}\widehat{r}_{(m,l)}=\sum_{(m,l)\in\mathcal{E}}r_{m,l}$. This scaling prevents large fluctuations in the value of the optimal regularization parameters empirically.
We refer to Problem (\ref{prop2}) as MTLACVX (\textbf{M}ulti-\textbf{T}ask \textbf{L}earning via \textbf{A}daptive \textbf{C}on\textbf{V}e\textbf{X} clustering). 
\subsection{Related work}

The proposed methods are related with some past studies (\cite{Zhou2011-xx}; \cite{Zhong2012-il}; \cite{Han2015-wp}). We describe the relationships and differences in this subsection.

For Problem (\ref{prop1}), we set a new variable $\bm{v}_{m} = \bm{w}_{m}-\bm{u}_{m}$. Then, the minimization problem is converted into the following minimization problem: 
    
    \begin{equation}
    \label{prop1ano}
        \min_{\substack{\bm{u}_{m},\bm{v}_{m}\in \mathbb{R}^{p},\\m=1,\ldots,T}}
        \left\{ \sum_{m=1}^{T} L(\bm{u}_{m}+\bm{v}_{m},w_{m0})+\frac{\lambda_{1}}{2}\sum_{m=1}^{T}\|\bm{v}_{m}\|_{2}^{2}+\lambda_{2}\sum_{(m,l)\in\mathcal{E}}r_{m,l}\|\bm{u}_{m}-\bm{u}_{l} \|_{2} \right\}.
    \end{equation}
    This minimization problem is regarded as an extension of Problem (\ref{groupregMTL}): it contains a multi-level structure for the regression coefficient vectors. This is close to \cite{Zhong2012-il}. However, they considered only using the $\ell_{1}$-norm for the fusion of $\bm{u}_{m}$ and the squared loss function. The $\ell_{1}$-norm penalty induces feature-level clustering rather than task-level clustering. On the other hand, they also proposed adapting weights for the fused penalty terms. The weights are calculated by using the estimated regression coefficient vectors $\widehat{\bm{w}}_{m}$, which may not be better for clustering than calculating the weights using $\widehat{\bm{u}}_{m}$, because $\widehat{\bm{w}}_{m}$ contains the value of $\widehat{\bm{v}}_{m}$.
    Moreover, they calculated adaptive weights for all of the combinations. Alternatively, we calculate adaptive weights $\widehat{r}_{m,l}$ only for $(m,l)\in\mathcal{E}$. 
    
    \cite{Han2015-wp} proposed MeTaG (Multi-Level Task Grouping) as follows:
    \begin{equation}
        \label{multilevel}
        \min_{\substack{\bm{w}_{m,h}\in \mathbb{R}^{p},\\m=1,\ldots,T,h=1,\ldots,H}}
        \left\{ \sum_{m=1}^{T} \frac{1}{2n_{m}}\|\bm{y}_{m} - \bm{X}_{m}\sum_{h=1}^{H}\bm{w}_{m,l} \|_{2}^{2}+\sum_{h=1}^{H}\lambda_{h}\sum_{m>l}\|\bm{w}_{m,h}-\bm{w}_{l,h} \|_{2} \right\},
    \end{equation}
where $\bm{w}_{m,h}\in\mathbb{R}^{p}$ is a parameter vector for $m$-th task and $h$-th level, $H$ is a total number of the level. In this minimization problem, the regression coefficient vector $\bm{w}_{m}$ is represented by the sum of the $h$-th level parameter vectors
as $\bm{w}_{m}=\sum_{h=1}^{H}\bm{w}_{m,h}$. Furthermore, each $h$-th level parameter is clustered by the second term. Because the aim of this minimization problem is not to improve the estimation accuracy for regression coefficient vectors and clustering but to capture complex multi-level structures, the proposed methods differ from this method in terms of their aim.

\cite{Zhou2011-xx} considered the following MTL method using $k$-means:
    \begin{equation}
        \label{k-meanMTL}
        \min_{\substack{\bm{w}_{m}\in \mathbb{R}^{p},m=1,\ldots,T,\\\mu_{c},\mathcal{I}_{c},c=1,\ldots,C}}
            \left\{ \sum_{m=1}^{T} \frac{1}{2n_{m}}\|\bm{y}_{m} - \bm{X}_{m}\bm{w}_{m} \|_{2}^{2}+\lambda \sum_{c=1}^{C}\sum_{m\in \mathcal{I}_{c}}\|\bm{w}_{m}-\bm{\mu}_{c} \|_{2}^{2}  \right\}, \\
    \end{equation}
    where $C$ is a number of cluster, $\bm{\mu}_{c}$ is a center of $c$-th cluster, and $\mathcal{I}_{c}$ is a set of task's index that belongs to $m$-th cluster.
   \cite{Zhou2011-xx} showed that \cite{Argyriou2007-cj} is a convex relaxation of Problem (\ref{k-meanMTL}). CVXMTL is also a convex relaxation of Problem (\ref{k-meanMTL}) in a different way.

\section{Estimation Algorithm}
\label{Sec4}

    In the proposed method, we compute the estimates of the parameters by the block coordinate descent algorithm (BCD).
  The BCD is performed by alternately computing the estimates: $\bm{u}_{m}$ is computed given $\bm{w}_{m}$, while $\bm{w}_{m}$ is done given $\bm{u}_{m}$. 

  We consider the two minimization problems:
     \begin{equation}
        \begin{split}
        \label{BCD}
           U^{(t+1)} &= \argmin_{\bm{u}_{m},m=1,\ldots,T}\left\{\frac{\lambda_{1}}{2}\sum_{m=1}^{T}\|\bm{w}_{m}^{(t)}-\bm{u}_{m}\|_{2}^{2}+\lambda_{2}\sum_{(m,l)\in\mathcal{E}}r_{m,l}\|\bm{u}_{m}-\bm{u}_{l} \|_{2}\right\},\\
        \bm{w}_{m}^{(t+1)} &= \argmin_{\bm{w}_{m},w_{m0}} \left\{  L(\bm{w}_{m},w_{m0})+ \frac{\lambda_{1}}{2}\|\bm{w}_{m}-\bm{u}_{m}^{(t+1)}\|_{2}^{2}\right\},\quad m=1,\ldots,T,
        \end{split}
    \end{equation}
    where superscript with brackets $(t)$ represents the number of updates and $U\in\mathbb{R}^{T\times p}$ is a matrix whose $m$-th row is $\bm{u}_{m}$. For the update of $\bm{w}_{m}$, when the loss function is a linear regression, it can be solved explicitly. When the logistic loss function is used, it can be solved 
 via the Newton-Raphson method, which is given by Algorithm \ref{Newton-Raphson}. In Algorithm \ref{Newton-Raphson}, we note that $\Pi$ is an $n\times n$-dimensional diagonal matrix, $\Lambda$ is a $p\times p$-dimensional diagonal matrix whose all diagonal elements are $\lambda_{1}$, $I_{n}$ is an $n\times n$-dimensional identity matrix, and $\bm{1}_{n}$ is an $n$-dimensional vector whose each element is one. 
 For the update of $\bm{u}_{m}$, we can compute it by using the algorithm for convex clustering such as \cite{Shimmura2022-ar} and \cite{Sun2021-jt}. In this paper, we adopt Algorithm \ref{CVXalgo} based on the idea of \cite{Shimmura2022-ar}, which converts the alternating direction method of multipliers (ADMM) into the proximal gradient method. This method enables us to use Nesterov's accelerated gradient method \citep{nesterov2003introductory} in the framework of ADMM. 
 
 As a result, the estimation algorithm for Problem (\ref{prop1}) is given by Algorithm \ref{Proposedalgo}. In Algorithm \ref{CVXalgo}, $A\in\mathbb{R}^{\abs{\mathcal{E}}\times T}, B\in\mathbb{R}^{T\times p}, W\in\mathbb{R}^{T\times p}, Z\in\mathbb{R}^{T\times p}$ and $S\in\mathbb{R}^{\abs{\mathcal{E}}\times p}$ are matrices whose each row components are constructed by $\bm{a}_{(m,l)}, \bm{b}_{m}, \bm{w}_{m}, \bm{z}_{m},$ and $\bm{s}_{(m,l)}$ vectors respectively. The vector $\bm{a}_{(m,l)}$ is defined as follows:
    \begin{equation}
        \label{definition_of_A}
        (\bm{a}_{(m,l)})_{j}=
        \begin{cases}
            1 \quad   &j=m,\\
            -1 \quad  &j=l,\\
            0 \quad &\mathrm{otherwise},
        \end{cases}
         j=1,\ldots,T.
    \end{equation}  
    $\mathrm{STL}(\cdot,\cdot)$ is a function returning an estimated regression coefficient vector by an arbitrary single-task learning method. $\mathrm{prox}(\cdot,\cdot)$ is defined as follows:
    \begin{equation}
     \mathrm{prox}(\bm{u},\lambda)=\mathrm{min}(\|\bm{u}\|_{2},\lambda)\frac{\bm{u}}{\|\bm{u}\|_{2}}.
    \end{equation}Because MTLCVX is a convex optimization problem and the BCD monotonically decreases the objective function, Algorithm \ref{Proposedalgo} converges to the global minimum. 
    See \cite{Shimmura2022-ar} for details of the Algorithm \ref{CVXalgo} and the way to determine the value of $\eta$.
   
    \begin{algorithm}[H]
    \caption{Estimation algorithm for Problem (\ref{prop1})}
    \label{Proposedalgo}
    \begin{algorithmic}
      \Require $\{\bm{y}_{m},X_{m}; m=1,\ldots,T\},k,\lambda_{1},\lambda_{2}$
      \For{ $m=1,\ldots,T$}
        \State $\widehat{\bm{w}}_{m}^{\mathrm{ST}}=  \mathrm{STL}(y_{m},X_{m})$
      \EndFor
      \State calculating $R$ by Eq. (\ref{setR}) from $k$ and $\widehat{\bm{w}}_{m}^{\mathrm{ST}}$
      \State $\bm{W}^{(0)}=\widehat{W}^{\mathrm{ST}}$
      \While{until convergence of $W^{(t)}$}
        \State $U^{(t+1)} = \mathrm{CVX}(W^{(t)},R,\lambda_{1},\lambda_{2})$
        \For{ $m=1\ldots,T$}
            \If{ $\bm{y}_{m}$ is a continuous response}
                \State $\bm{w}_{m}^{(t+1)}  = (\frac{1}{n_{m}}X_{m}^{\top}X_{m}+\lambda_{1}I_{p})^{-1}(\frac{1}{n_{m}}X_{m}^{\top}\bm{y}_{m}+\lambda_{1}\bm{u}_{m}^{(t+1)})$
            \EndIf
            \If{$\bm{y}_{m}$ is a binary response}
                \State $(w_{m0}^{(t+1)},\bm{w}_{m}^{(t+1)\top})^{\top}  = \mathrm{NR}(n_{m},X_{m},\bm{y}_{m},\bm{u}_{m}^{(t+1)},\lambda_{1})$
            \EndIf
        \EndFor
       \EndWhile
      \Ensure $U,W,w_{m0},m=1,\ldots,T$
    \end{algorithmic}
  \end{algorithm}
  
  \begin{algorithm}[H]
    \caption{Estimation algorithm for the convex clustering}
    \label{CVXalgo}
    \begin{algorithmic}
      \Function {CVX}{$W,R,\lambda_{1},\lambda_{2}$}
      \State Initialize; $U^{(0)}=W$
      \State converting $R$ into $A$ by Eq (\ref{definition_of_A})
      \State $G=A^{\top}A,\eta=\frac{1}{\lambda_{1}+2\max_{i=1,\ldots,T}((G)_{ii})}$
      \While{until convergence of $U^{(t)}$}
     \State $k=0,\alpha^{(0)} =1, Z^{(0)}=U^{(t)},B^{(1)}=U^{(t)}$
      \While{until convergence of $Z^{(k)}$}
        \State $D=S^{(t)}+\rho A \cdot B^{(k)}$
        \For{ $(m,l)\in\mathcal{E}$}
            \State $D^{\mathrm{prox}}_{(m,l)}=\mathrm{prox}(\bm{d}_{(m,l)},\lambda_{2}r_{m,l})$
         \EndFor
        \State $Z^{(k+1)} = B^{(k)}-\eta\{\lambda_{1}(B^{(k)}-W)+A^{\top}D^{\mathrm{prox}}\} $
        \State $\alpha^{(k+1)}=\frac{1+\sqrt{1+4(\alpha^{(k)})^{2}}}{2}$
        \State $B^{(k+1)} =B^{(k)}+\frac{\alpha^{(k)}-1}{\alpha^{(k+1)}}(Z^{(k)}-Z^{(k-1)})$
      \EndWhile
       \State $U^{(t+1)}=Z^{(k)}$
         \State $S^{\prime}=S^{(t)}+\rho A \cdot U^{(t+1)}$
         \For{$(m,l)\in\mathcal{E}$}
            \State $S^{(t+1)}_{(m,l)} = \mathrm{prox}(\bm{s}_{(m,l)}^{\prime},\lambda_{2}r_{m,l})$
         \EndFor
      \EndWhile
      \State Output: $U$
    \EndFunction
    \end{algorithmic}
  \end{algorithm}

  \begin{algorithm}[H]
    \caption{Newton-Raphson method for solving logistic regression}
    \label{Newton-Raphson}
    \begin{algorithmic}
      \Function {NR}{$n,X,\bm{y},\bm{u},\lambda_{1}$}
            \State Initialize; $w_{0}^{(t)}=0,\bm{w}^{(k)}=\bm{0}$
      \While{until convergence of $w_{0},\bm{w}$}
        \State $(\Pi^{(t+1)})_{ii} = 1-1/(1+\exp(w_{0}^{(t)}+\bm{w}^{(t)\top} \bm{x}_{i})$ for $i=1,\ldots,n.$
        \State $w_{0}^{(t+1)} = w_{0}^{(t)} + (\bm{1}_{n}^{\top} \Pi^{(t)}(I_{n}-\Pi^{(t)}) \bm{1}_{n})^{-1}(\bm{1}_{n}^{\top}(\bm{y}-\Pi^{(t)} \bm{1}_{n}))$
        \State $\bm{w}^{(t+1)} = \bm{w}^{(t)} + (X^{\top} \Pi^{(t)}(I_{n}-\Pi^{(t)})X/n +\Lambda)^{-1}(X^{\top}(\bm{y}-\Pi^{(t)} \bm{1}_{n})/n-\Lambda(\bm{w}^{(t)}-\bm{u}))$        
      \EndWhile
      \State Output: $(w_{0},\bm{w}^{\top})^{\top}$
    \EndFunction
    \end{algorithmic}
  \end{algorithm}

\section{Simulation studies}

\label{Sec5}
In this section, we report simulation studies in the linear regression setting. We have generated data by the true model:
    \begin{equation}
        \begin{split}
            \bm{y}_{m} = X_{m}\bm{w}_{m}^{\ast}+\bm{\epsilon}_{m}, \quad m=1,\ldots,T,
        \end{split}
    \end{equation}
where $\bm{\epsilon}_{m}$ is an error term whose each component is distributed as $N(0,\sigma^{2})$ independently, $\bm{w}^{\ast}_{m}$ is a true regression coefficient vector for $m$-th task. For this true model, these $T$ tasks consist of $C$ true clusters. The number of tasks in each cluster is uniformly set by $T/C$.
The design matrix $X_{m}$ is generated from $N_{p}(\bm{0},\Sigma)$ for each task independently, where $(\Sigma)_{ij} = \phi^{\abs{i-j}}$.

 The true regression coefficient vector $\bm{w}_{m}^{\ast}$ is generated as follows. 
 First, each explanatory variable $\{j=1,\ldots,p\}$ was randomly assigned to the $c$-th clusters $\{c=1,\ldots,C \}$ with the same probability. Then, we generated a true centroid parameter for $c$-th cluster $\bm{u}^{\ast}_{c} = (u^{\ast}_{c1},\ldots,u^{\ast}_{cp})^{\top}$ by
 \begin{equation}
  u_{cj}^{\ast}
  \begin{cases}
    \sim N(0,\sigma_{u}^{2}) & \text{if $j$-th variable is assigned to $c$-th cluster},\\
    =0 & \text{otherwise},
  \end{cases}
    \quad j=1,\ldots,p.
 \end{equation}
 In addition, we generated a true task-specific parameter for $m$-th task 
that belongs to $c$-th cluster $\bm{v}^{(c)\ast}_{m} = (v^{(c)\ast}_{m1},\ldots,v^{(c)\ast}_{mp})^{\top}$ by
 \begin{equation}
  v_{mj}^{(c)\ast}
  \begin{cases}
    \sim N(0,\sigma_{v}^{2}) & \text{if $j$-th variable is assigned to $c$-th cluster},\\
    =0 & \text{otherwise},
  \end{cases}
      \quad j=1,\ldots,p.
 \end{equation}
Finally, we set to $\bm{w}_{m}^{\ast}=\bm{u}_{c}^{\ast} + \bm{v}_{m}^{(c)\ast}$. In this way, regression coefficient vectors belonging to different clusters have different non-zero variables. A similar way of generating regression coefficient vectors was also used in \cite{Zhou2016-ik}.

 For our true model, we set setting as $n_{m}=230$, $p=100$, $T=100$, $\sigma^{2}=5$ and $\sigma_{u}^{2} = 100$. $230$ samples in each task were split into $30$ samples for the train, $100$ samples for the validation, and left samples for the test.  We  considered  several  settings: $\phi=\{0, 0.2, 0.5\}$, $\sigma_{v}^{2}=\{1,2,3,4,5\}$, and  $C=\{5, 10\}$. 
 
 To evaluate the effectiveness of our proposed methods, we compared them with the single-task learned lasso (STLL) and the multi-task learning via network lasso (MTLNL). STLL is conducted by estimating each task by the lasso independently. MTLNL is Problem (\ref{groupregMTL}) for $q=2$, which is estimated by ADMM. The weights $r_{m,l}$ for both MTLNL and MTLCVX were calculated by Eq. (\ref{setR}). In this case, $k$ was set to five. The estimation of both STLL and $\widehat{\bm{w}}_{m}^{\mathrm{SL}}$ in Eq. (\ref{setR}) were performed by the lasso in R package ``glmnet''. The tuning parameter $\rho$ included in Algorithm \ref{Proposedalgo} and ADMM to estimate MTLNL were set to one. The regularization parameters except for STLL were determined by the validation data. For the evaluation, 
we calculated the NMSE (normalized mean squared error) and RMSE (root mean squared error) as follows:
    \begin{equation}
        \begin{split}
        \label{evaluation}
        \mathrm{NMSE} &= \frac{1}{T} \sum_{m=1}^{T}\frac{\|\bm{y}^{\ast}_{m}-X_{m}\widehat{\bm{w}}_{m} \|_{2}^{2}}{\mathrm{Var}(\bm{y_{m}^{\ast}})},\\
        \mathrm{RMSE} &= \frac{1}{T}\sum_{m=1}^{T}\|\bm{w}^{\ast}_{m}-\widehat{\bm{w}}_{m} \|_{2}.
        \end{split}
    \end{equation}
These values evaluate the accuracy of the prediction and estimated regression coefficient vectors, respectively. They were computed 100 times.  The mean and standard deviation were obtained in each setting.

Tables \ref{simresult1} and \ref{simresult2} show the results of the simulation studies for $C=10$ and $C=5$, respectively. Since STLL is independent of the value of $\sigma_{v}^{2}$, we show the results for STLL only when $\sigma_{v}^{2}=1$.
Note that, according to decreasing the value of $C$, the number of the true non-zero variables in each task is increased, because variables are nonzero only in the cluster to which they are assigned. Then, the results of STLL in Table \ref{simresult2} considerably deteriorate. This also indicates that the weights $r_{m,l}$ contain more noise at $C=5$ than at $C=10$. Thus, the results of Table \ref{simresult2} are worse than Table \ref{simresult1} on the whole.

In a comparison among the methods, MTLACVX shows superior accuracy in almost all situations for both NMSE and RMSE. The differences between MTLACVX and MTLCVX or MTLNL are much larger than that between MTLCVX and MTLNL.  Thus, in the context of convex clustering, it means that the adaptive weights are important for improving estimation accuracy. On the other hand, for the comparison of MTLNL and MTLCVX, MTLCVX shows better performance than MTLNL on the whole. In particular, when $C=5$, MTLCVX is superior to MTLNL in all settings except for NMSE in $\phi=0$ and $\sigma_{v}^{2}=5$. When $C=10$, again, MTLCVX is superior to MTLNL in many settings. MTLNL shows better results than MTLCVX for two settings only when $\phi=0$. It probably relates the estimation accuracy of $\widehat{\bm{w}}_{m}^\mathrm{(SL)}$ to construct wetghts $r_{m,l}$ by Eq. (\ref{setR}). For STLL, RMSE drastically deteriorates by increasing the value of $\phi$ from $0$ to $0.2$. This also indicates that the noise in weights $r_{m,l}$ also increased from $\phi=0$ to $\phi = 0.2$. Hence, there is not much difference between MTLNL and MTLCVX for $\phi=0$, because there was less noise in the weights. However, MTLCVX would be superior to MTLCVX as the noise in the weights increased. On the whole, these results suggest that MTLCVX is more robust to the noise in the weights $r_{m,l}$ than MTLNL.
    \begin{table}[!ht]
    \centering
    \caption{Mean and standard deviation of NMSE and RMSE for $C=10$}
    \label{simresult1}
    \scalebox{0.8}
    {
    \begin{tabular}{cc|cc|cc|cc}
    \hline
         &  & \multicolumn{2}{c}{$\phi = 0$} &   \multicolumn{2}{c}{$\phi = 0.2$} & \multicolumn{2}{c}{$\phi = 0.5$}   \\ \hline
        $\sigma_{v}^2$ & method & NMSE & RMSE & NMSE & RMSE & NMSE & RMSE  \\ \hline
        - & STLL & 0.200 (0.025) & 1.494 (0.161) & 0.198 (0.034) & 4.062 (0.284) & 0.178 (0.033) & 4.070( 0.315)  \\ \hline
        $1$ & MTLNL & 0.059 (0.028) & 0.646 (0.146) & 0.053 (0.032) & 0.614 (0.148) &  0.049 (0.029) & 0.664 (0.176)  \\
        ~ & MTLCVX & 0.055 (0.038) & 0.609 (0.190) & 0.048 (0.023) & 0.596 (0.155) & 0.039 (0.018) &  0.574 (0.147)  \\
        ~ & MTLACVX & $\bm{0.044}$ (0.019) & $\bm{0.565}$ (0.158) & $\bm{0.043}$ (0.029) & $\bm{0.559}$ (0.152) & $\bm{0.038}$ (0.023) & $\bm{0.565}$ (0.156)  \\ \hline
        $2$ & MTLNL & 0.075 (0.029) & 0.741 (0.127) & 0.068 (0.024)  & 0.755 (0.163) & 0.063 (0.030) & 0.762 (0.137)  \\ 
        ~ & MTLCVX & 0.063 (0.026) & $\bm{0.689}$ (0.148) & 0.058 (0.023) & 0.667 (0.151) & 0.052 (0.021) & 0.691 (0.138)  \\ 
        ~ & MTLACVX & $\bm{0.060}$ (0.024) & 0.696 (0.185) & $\bm{0.055}$ (0.020) & $\bm{0.637}$ (0.125) & $\bm{0.048}$ (0.020) & $\bm{0.666}$ (0.124)  \\  \hline
        $3$ & MTLNL & 0.083 (0.049) & 0.789 (0.160) & 0.080 (0.023) & 0.815 (0.130) & 0.076 (0.036) &  0.868 (0.141)  \\ 
        ~ & MTLCVX & $\bm{0.080}$ (0.035) & 0.775 (0.157) & 0.078 (0.038) & 0.767 (0.119) & 0.066 (0.021) & 0.791 (0.126)  \\ 
        ~ & MTLACVX & 0.081 (0.035) & $\bm{0.771}$ (0.138) &  $\bm{0.073}$ (0.024) & $\bm{0.752}$ (0.124) & $\bm{0.065}$ (0.027) & $\bm{0.764}$ (0.122)  \\ \hline
        $4$ & NLMTL & 0.106 (0.077) & 0.906 (0.130) & 0.093 (0.026) & 0.889 (0.138) & 0.079 (0.020) & 0.906 (0.090)  \\ 
        ~ & MTLCVX & $\bm{0.084}$ (0.027) & 0.818 (0.111) & 0.090 (0.033) & 0.856 (0.122) & 0.076 (0.024) & 0.861 (0.123)  \\ 
        ~ & MTLACVX & 0.085 (0.025) & $\bm{0.815}$ (0.126) & $\bm{0.084}$ (0.024) & $\bm{0.831}$ (0.129) & $\bm{0.074}$ (0.024) & $\bm{0.841}$ (0.112)  \\ \hline
        $5$ & MTLNL & 0.105 (0.029) & 0.921 (0.099) & 0.102 (0.025) & 0.939 (0.111) & 0.096 (0.043) & 0.998 (0.118)  \\ 
        ~ & MTLCVX & 0.113 (0.099) & 0.906 (0.129) & 0.099 (0.030) & 0.898 (0.124) & 0.088 (0.025) & 0.938 (0.109)  \\ 
        ~ & MTLACVX & $\bm{0.099}$ (0.032) & $\bm{0.888}$ (0.120) & $\bm{0.094}$ (0.025) & $\bm{0.894}$ (0.130) & $\bm{0.087}$ (0.027) & $\bm{0.931}$ (0.114)  \\ \hline
    \end{tabular}
    }
\end{table}

\begin{table}[!ht]
    \centering
    \caption{Mean and standard deviation of NMSE and RMSE for $C=5$}
    \label{simresult2}
    \scalebox{0.8}
    {
    \begin{tabular}{cc|cc|cc|cc}
    \hline
         &   & \multicolumn{2}{c}{$\phi = 0$} & \multicolumn{2}{c}{$\phi = 0.2$} & \multicolumn{2}{c}{$\phi = 0.5$} \\ \hline
        $\sigma_{v}^{2}$ & method & NMSE & RMSE & NMSE & RMSE & NMSE & RMSE \\ \hline
        - & STLL & 0.567 (0.046) & 3.423 (0.284) & 0.564 (0.054) &  5.241 (0.055) & 0.493 (0.054) & 5.395(0.359) \\ \hline
        $1$ & MTLNL & 0.127 (0.059) & 1.543 (0.394) & 0.117 (0.055) & 1.475 (0.388) & 0.080 (0.048) & 1.337 (0.396) \\ 
        ~ & MTLCVX & 0.131 (0.055) & 1.559 (0.384) & 0.113 ( 0.055) & 1.459 (0.379) & 0.074 (0.041) & 1.291 (0.393) \\ 
        ~ & MTLACVX & $\bm{0.105}$ (0.053) & $\bm{1.361}$ (0.430) & $\bm{0.112}$ (0.056) & $\bm{1.415}$ (0.403) & $\bm{0.070}$ (0.039) & $\bm{1.225}$ (0.372) \\ \hline
        $2$ & MTLNL & 0.145 (0.054) & 1.625 (0.350) & 0.139 (0.055) & 1.627 (0.373) &  0.086 (0.043) & 1.435 (0.377) \\ 
        ~ & MTLCVX & 0.142 (0.058) & 1.606 (0.339) & 0.131 (0.045) & 1.582 (0.318) & 0.078 (0.043) & 1.296 (0.352) \\ 
        ~ & MTLACVX & $\bm{0.132}$ (0.062) & $\bm{1.557}$ (0.423) & $\bm{0.112}$ (0.056) & $\bm{1.470}$ (0.380) & $\bm{0.078}$ (0.041) & $\bm{1.293}$ (0.336) \\ \hline
        $3$ & MTLNL & 0.151 (0.056) & 1.689 (0.339) & 0.146 (0.051) & 1.710 (0.334) & 0.102 (0.035) & 1.545 (0.308) \\ 
        ~ & MTLCVX & 0.159 (0.054) & 1.730 (0.339) & 0.134 (0.061)  & 1.582 (0.384) & $\bm{0.090}$ (0.038) & $\bm{1.442}$ (0.297) \\ 
        ~ & MTLACVX & $\bm{0.132}$ (0.054) & $\bm{1.532}$ (0.369) & $\bm{0.119}$ (0.049) & $\bm{1.495}$ (0.316) & 0.094 (0.045) & 1.444 (0.373) \\ \hline 
        $4$ & MTLNL & 0.162 ( 0.054) & 1.774 (0.339) & 0.162 (0.057) & 1.801 (0.344) & 0.108 (0.040) & 1.575 (0.287) \\ 
        ~ & MTLCVX & 0.155 (0.060) & 1.716 (0.351) & 0.154 (0.056) & 1.746 (0.340) & 0.099 (0.041) & 1.533 (0.340) \\ 
        ~ & MTLACVX & $\bm{0.145}$ (0.059) & $\bm{1.667}$ (0.397) & $\bm{0.120}$ (0.044) & $\bm{1.515}$ (0.309) & $\bm{0.094}$ (0.044) & $\bm{1.460}$ (0.364) \\ \hline
        $5$ & MTLNL & 0.179 (0.061) & 1.853 (0.330) & 0.169 (0.051) & 1.850 (0.303) & 0.117 (0.040) & 1.669 (0.298) \\ 
        ~ & MTLCVX & 0.163 (0.057) & 1.757 (0.315) & 0.157 (0.051) & 1.773 (0.311) & 0.119 (0.046) & 1.674 (0.323) \\ 
        ~ & MTLACVX & $\bm{0.146}$ (0.041) & $\bm{1.706}$ (0.274) & $\bm{0.130}$ (0.042) & $\bm{1.612}$ (0.276) & $\bm{0.093}$ (0.036) &  $\bm{1.481}$ (0.288)  \\ \hline
    \end{tabular}
    }
\end{table}

\section{Application to real datasets}
\label{Sec6}

In this section, we applied our proposed methods to two datasets with continuous and binary responses. 
The first is the school data \citep{Bakker2003-tp}, which has been often used as the research of an MTL. This dataset consists of examination scores of 15,362 students, school-specific attributes, and student-specific attributes from 139 secondary schools in London from 1985 to 1987. The examination scores are used as a response and other features as 27-dimensional explanatory variables. Each school is considered as a task. The second is the landmine data \citep{Xue2007-wl}, which consists of nine-dimensional features and the corresponding binary labels for 29 tasks. The responses represent landmines or clutter. Though there are 14,820 samples in total, this dataset is quite unbalanced: positive samples are few, while negative ones are many. To perform our proposed method, down-sampling was done by reducing negative samples to equal the number of positive samples. In the results, we used 1,808 samples in total. 

 We compared our proposed methods MTLCVX, MTLACVX with MTLNL, STLL, and single-task learned ridge (STLR), where STLR is the ridge estimation performed by R package ``glmnet'' for each task, independently. Note that, to stabilize estimation in the logistic regression of MTLNL, MTLCVX, and MTLACVX, we penalized the intercept $w_{m0}$ by the ridge. Its regularization parameter was set to 0.1. This penalty has the effect of keeping the intercept constant finite stable value in the situation that the intercept tends to go to infinity. We randomly split the data into $V\%$ of the data for the train, $(80-V)\%$ for the test, and $20\%$ for the validation. We  conducted three settings $V=\{50,60,70\}$. For the evaluation, we used NMSE for analyzing the school data, while we used AUC for analyzing the landmine data. The mean and standard deviation of evaluation values were computed from 100 repetitions.
 The tuning parameter $k$ in Eq. (\ref{setR}) was set to five for all MTL methods and $\widehat{\bm{w}}_{m}^{\mathrm{SL}}$ were estimated by the lasso by the package ``glmnet" in R.
 
 Table \ref{real_linear} shows the results of the school data for each setting. First, all MTL methods are superior to single-task learning approaches. In a comparison among MTL methods, each method shows a better result for each setting. However, because all settings have outstanding standard deviations for $V=70$, this result is probably not trustworthy. Though the school data is often used in the research of an MTL, \cite{Evgeniou2005-it} pointed out that the data do not have clusters and are rather homogenous. Therefore, the data may be more favorable to MTLNL, because MTLNL is more likely to shrink the difference in tasks.
 
Table \ref{real_logi} shows the results of the landmine data for each setting. In the data, MTLACVX and MTLCVX are superior to STL methods and MTLNL for all settings. MTLACVX also has the same or better performance than MTLCVX. Unlike the school data, the landmine data is considered to have two clusters:  highly foliated regions and bare earth or desert regions. Hence, the data has more distinct clusters than the school data. This may be the reason that MTLCVX and MTLACVX in the landmine data provide higher accuracy compared to those in the school data. 
\begin{table}[!ht]
    \caption{Mean and standard deviation of NMSE for 100 repetitions in the school data}
    \label{real_linear}
    \centering
    \begin{tabular}{cccc}
        \hline
        method & 50\% & 60\% & 70\%   \\ \hline
        STLL &  4.044 (0.181) & 4.293 (0.234) & 5.783 (1.306) \\ 
        STLR &  4.701 (0.226) & 5.071 (0.516) & 6.533 (1.170) \\ 
        MTLNL & 0.806 (0.025) & $\bm{0.847}$ (0.036) & 1.196 (0.517) \\ 
        MTLCVX &  $\bm{0.796}$ (0.025) & 0.853 (0.060) & 1.241 (0.825) \\ 
        MTLACVX & 0.830 (0.036) & 0.863 (0.060) & $\bm{1.140}$ (0.528) \\ \hline
    \end{tabular}
\end{table}

\begin{table}[!ht]
    \centering
    \caption{Mean and standard deviation of AUC for 100 repetitions in the landmine data}
        \label{real_logi}
    \begin{tabular}{cccc}
        \hline
        method & 50\% & 60\% & 70\%  \\ \hline
        STLL & 0.746 (0.023) & 0.748 (0.022) & 0.748 (0.020)  \\ 
        STLR & 0.749 (0.023) & 0.750 (0.023) & 0.749 (0.027)  \\ 
        MTLNL  & 0.754 (0.024) & 0.749 (0.024) & 0.750 (0.021)  \\ 
        MTLCVX & $\bm{0.769}$ (0.021) & 0.759 (0.020) & 0.760 (0.023)  \\ 
        MTLACVX & 0.768 (0.018) & $\bm{0.764}$ (0.022) & $\bm{0.770}$ (0.023)  \\ 
        \hline
    \end{tabular}
\end{table}

\section{Conclusion}
\label{Sec7}

In this paper, we proposed the MTL method referred to as MTLCVX.
Because the parameters are split into those for regression and for clustering, we can expect to reduce the shrinkages between irrelevant tasks, which is caused by fused group regularization.
In simulation studies, our proposed methods show better results compared with the existing method by the network lasso in almost all cases. MTLCVX can be more robust against noise in the weights than MTLNL. For the application to real data, if there are distinct cluster structures in the data, MTLCVX shows better performance.

 We can also extend the proposed method based on the research of \cite{Wang2018-xf} and \cite{Quan2020-mc}. For example, sparse convex clustering could be introduced to reduce the number of features used for clustering. Also, robust convex clustering could be introduced to exclude outlier tasks from the cluster. These extensions would be easily implemented by replacing Algorithm \ref{CVXalgo} with their estimation algorithm. On the other hand, although our study used a $k$-nearest neighbor to construct weights, there may be better methods in terms of both computational complexity and estimation accuracy. We leave this topic as future work.
\backmatter
\bmhead{Acknowledgments}
S. K. was supported by JSPS KAKENHI Grant Numbers JP19K11854 and JP23K11008. Supercomputing resources were provided by the Human Genome Center (the Univ. of Tokyo).
\noindent

\bigskip




\bibliography{CVXMTL.bib}

\begin{thebibliography}{34}
\providecommand{\natexlab}[1]{#1}
\providecommand{\url}[1]{\texttt{#1}}
\expandafter\ifx\csname urlstyle\endcsname\relax
  \providecommand{\doi}[1]{doi: #1}\else
  \providecommand{\doi}{doi: \begingroup \urlstyle{rm}\Url}\fi

\bibitem[Ando and Zhang(2005)]{Ando2005-pl}
Ando, R.~K. and Zhang, T.
\newblock (2005).
\newblock A framework for learning predictive structures from multiple tasks
  and unlabeled data.
\newblock \emph{Journal of Machine Learning Research}, {\bfseries 6}, \penalty0
  1817--1853.

\bibitem[Argyriou et~al.(2007)Argyriou, Pontil, Ying, and
  Micchelli]{Argyriou2007-cj}
Argyriou, A., Pontil, M., Ying, Y., and Micchelli, C.~A.
\newblock (2007).
\newblock A spectral regularization framework for multi-task structure
  learning.
\newblock \emph{Advances in Neural Information Processing Systems}, {\bfseries
  20}, \penalty0 25--32.

\bibitem[Bakker and Heskes(2003)]{Bakker2003-tp}
Bakker, B. and Heskes, T.
\newblock (2003).
\newblock Task clustering and gating for bayesian multitask learning.
\newblock \emph{Journal of Machine Learning Research}, {\bfseries 4}, \penalty0
  83--99.

\bibitem[Boyd et~al.(2011)Boyd, Parikh, Chu, Peleato, Eckstein,
  et~al.]{Boyd2011-ul}
Boyd, S., Parikh, N., Chu, E., Peleato, B., Eckstein, J., et~al.
\newblock (2011).
\newblock Distributed optimization and statistical learning via the alternating
  direction method of multipliers.
\newblock \emph{Foundations and Trends{\textregistered} in Machine learning},
  {\bfseries 3}\penalty0 (1), \penalty0 1--122.

\bibitem[Caruana(1997)]{Caruana1997-bb}
Caruana, R.July .
\newblock (1997).
\newblock Multitask learning.
\newblock \emph{Machine learning}, {\bfseries 28}, \penalty0 41--75.

\bibitem[Deng et~al.(2017)Deng, Shahabi, Demiryurek, and Zhu]{Deng2017-oc}
Deng, D., Shahabi, C., Demiryurek, U., and Zhu, L.
\newblock (2017).
\newblock Situation aware multi-task learning for traffic prediction.
\newblock In \emph{2017 IEEE International Conference on Data Mining},  81--90.

\bibitem[Dondelinger et~al.(2020)Dondelinger, Mukherjee, and {Alzheimer's
  Disease Neuroimaging Initiative}]{Dondelinger2020-wz}
Dondelinger, F., Mukherjee, S., and {Alzheimer's Disease Neuroimaging
  Initiative}.
\newblock (2020).
\newblock The joint lasso: high-dimensional regression for group structured
  data.
\newblock \emph{Biostatistics}, {\bfseries 21}\penalty0 (2), \penalty0
  219--235.

\bibitem[Evgeniou et~al.(2005)Evgeniou, Micchelli, and Pontil]{Evgeniou2005-it}
Evgeniou, T., Micchelli, C.~A., and Pontil, M.
\newblock (2005).
\newblock Learning multiple tasks with kernel methods.
\newblock \emph{Journal of Machine Learning Research}, {\bfseries 6}, \penalty0
  615--637.

\bibitem[Fan et~al.(2008)Fan, Gao, and Luo]{Fan2008-ad}
Fan, J., Gao, Y., and Luo, H.
\newblock (2008).
\newblock Integrating concept ontology and multitask learning to achieve more
  effective classifier training for multilevel image annotation.
\newblock \emph{IEEE Transactions on Image Processing}, {\bfseries 17}\penalty0
  (3), \penalty0 407--426.

\bibitem[Hallac et~al.(2015)Hallac, Leskovec, and Boyd]{Hallac2015-ss}
Hallac, D., Leskovec, J., and Boyd, S.
\newblock (2015).
\newblock Network lasso: Clustering and optimization in large graphs.
\newblock In \emph{Proceedings of the 21th ACM SIGKDD International Conference
  on Knowledge Discovery and Data Mining},  387--396.

\bibitem[Han and Zhang(2015)]{Han2015-wp}
Han, L. and Zhang, Y.
\newblock (2015).
\newblock Learning multi-level task groups in multi-task learning.
\newblock In \emph{Proceedings of the AAAI Conference on Artificial
  Intelligence}, {\bfseries 29}\penalty0 (1), \penalty0 2638--2644.

\bibitem[He et~al.(2019)He, Alesiani, and Shaker]{He2019-wy}
He, X., Alesiani, F., and Shaker, A.
\newblock (2019).
\newblock Efficient and scalable multi-task regression on massive number of
  tasks.
\newblock In \emph{Proceedings of the AAAI Conference on Artificial
  Intelligence}, {\bfseries 33}\penalty0 (01), \penalty0 3763--3770.

\bibitem[Hocking et~al.(2011)Hocking, Joulin, Bach, and Vert]{Hocking2011-mn}
Hocking, T.~D., Joulin, A., Bach, F., and Vert, J.~P.
\newblock (2011).
\newblock Clusterpath an algorithm for clustering using convex fusion
  penalties.
\newblock In \emph{Proceedings of the 28th International Conference on Machine
  Learning},  745--752.

\bibitem[Kang et~al.(2011)Kang, Grauman, and Sha]{Kang2011-gs}
Kang, Z., Grauman, K., and Sha, F.
\newblock (2011).
\newblock Learning with whom to share in multi-task feature learning.
\newblock In \emph{Proceedings of the 28th International Conference on Machine
  Learning},  521--528.

\bibitem[Li et~al.(2018)Li, He, and Borgwardt]{Li2018-tf}
Li, L., He, X., and Borgwardt, K.
\newblock (2018).
\newblock Multi-target drug repositioning by bipartite block-wise sparse
  multi-task learning.
\newblock \emph{BMC Systems Biology}, {\bfseries 12}\penalty0 (4), \penalty0
  85--97.

\bibitem[Lindsten et~al.(2011)Lindsten, Ohlsson, and Ljung]{Lindsten2011-bs}
Lindsten, F., Ohlsson, H., and Ljung, L.
\newblock (2011).
\newblock Clustering using sum-of-norms regularization: With application to
  particle filter output computation.
\newblock In \emph{2011 IEEE Statistical Signal Processing Workshop},
  201--204.

\bibitem[Nesterov()]{nesterov2003introductory}
Nesterov, Y.
\newblock Introductory lectures on convex optimization: A basic course.
\newblock {\bfseries 87}, \penalty0 1--78.

\bibitem[Obozinski et~al.(2010)Obozinski, Taskar, and Jordan]{Obozinski2010-rt}
Obozinski, G., Taskar, B., and Jordan, M.~I.
\newblock (2010).
\newblock Joint covariate selection and joint subspace selection for multiple
  classification problems.
\newblock \emph{Statistics and Computing}, {\bfseries 20}, \penalty0 231--252.

\bibitem[Parameswaran and Weinberger()]{Parameswaran2010-of}
Parameswaran, S. and Weinberger, K.~Q.
\newblock Large margin multi-task metric learning.
\newblock \emph{Advances in Neural Information Processing Systems}, {\bfseries
  23}, \penalty0 1867--1875.

\bibitem[Pelckmans et~al.(2005)Pelckmans, De~Brabanter, Suykens, and
  De~Moor]{Pelckmans2005-kw}
Pelckmans, K., De~Brabanter, J., Suykens, J., and De~Moor, B.
\newblock (2005).
\newblock Convex clustering shrinkage.
\newblock In \emph{PASCAL workshop on Statistics and Optimization of Clustering
  workshop}.

\bibitem[Quan and Chen(2020)]{Quan2020-mc}
Quan, Z. and Chen, S.Jan. .
\newblock (2020).
\newblock Robust convex clustering.
\newblock \emph{Soft computing}, {\bfseries 24}\penalty0 (2), \penalty0
  731--744.

\bibitem[Shimamura and Kawano(2021)]{Shimamura2021-ma}
Shimamura, K. and Kawano, S.
\newblock (2021).
\newblock A bayesian approach to multi-task learning with network lasso.
\newblock Preprint, arXiv:1402.6455.

\bibitem[Shimmura and Suzuki(2022)]{Shimmura2022-ar}
Shimmura, R. and Suzuki, J.
\newblock (2022).
\newblock Converting admm to a proximal gradient for efficient sparse
  estimation.
\newblock \emph{Japanese Journal of Statistics and Data Science}, (Online
  Access).

\bibitem[Sun et~al.(2021)Sun, Toh, and Yuan]{Sun2021-jt}
Sun, D., Toh, K.-C., and Yuan, Y.
\newblock (2021).
\newblock Convex clustering: Model, theoretical guarantee and efficient
  algorithm.
\newblock \emph{Journal of Machine Learning Research}, {\bfseries 22}\penalty0
  (1), \penalty0 427--458.

\bibitem[Tan and Witten(2015)]{Tan2015-oo}
Tan, K.~M. and Witten, D.
\newblock (2015).
\newblock Statistical properties of convex clustering.
\newblock \emph{Electronic Journal of Statistics}, {\bfseries 9}\penalty0 (2),
  \penalty0 2324--2347.

\bibitem[Wang et~al.(2018)Wang, Zhang, Sun, and Fang]{Wang2018-xf}
Wang, B., Zhang, Y., Sun, W.~W., and Fang, Y.
\newblock (2018).
\newblock Sparse convex clustering.
\newblock \emph{Journal of Computational and Graphical Statistics}, {\bfseries
  27}\penalty0 (2), \penalty0 393--403.

\bibitem[Xue et~al.(2007)Xue, Liao, Carin, and Krishnapuram]{Xue2007-wl}
Xue, Y., Liao, X., Carin, L., and Krishnapuram, B.
\newblock (2007).
\newblock Multi-task learning for classification with dirichlet process priors.
\newblock \emph{Journal of Machine Learning Research}, {\bfseries 8}, \penalty0
  35--63.

\bibitem[Yamada et~al.(2017)Yamada, Koh, Iwata, Shawe-Taylor, and
  Kaski]{Yamada2017-cg}
Yamada, M., Koh, T., Iwata, T., Shawe-Taylor, J., and Kaski, S.
\newblock (2017).
\newblock {Localized Lasso for High-Dimensional Regression}.
\newblock In \emph{Proceedings of the 20th International Conference on
  Artificial Intelligence and Statistics}, {\bfseries 54}, \penalty0 325--333.

\bibitem[Zhang et~al.(2022)Zhang, Liu, and Zhu]{Zhang2022-lt}
Zhang, X., Liu, J., and Zhu, Z.
\newblock (2022).
\newblock Learning coefficient heterogeneity over networks: A distributed
  spanning-tree-based fused-lasso regression.
\newblock \emph{Journal of the American Statistical Association}, (early
  access).

\bibitem[Zhong and Kwok(2012)]{Zhong2012-il}
Zhong, W. and Kwok, J. T.~Y.
\newblock (2012).
\newblock Convex multitask learning with flexible task clusters.
\newblock In \emph{Proceedings of the 29th International Conference on Machine
  Learning ICML 2012},  49--56.

\bibitem[Zhou et~al.(2011{\natexlab{a}})Zhou, Chen, and Ye]{Zhou2011-xx}
Zhou, J., Chen, J., and Ye, J.
\newblock (2011).
\newblock Clustered multi-task learning via alternating structure optimization.
\newblock \emph{Advances in Neural Information Processing Systems}, {\bfseries
  24}, \penalty0 702--710.

\bibitem[Zhou et~al.(2011{\natexlab{b}})Zhou, Yuan, Liu, and Ye]{Zhou2011-wm}
Zhou, J., Yuan, L., Liu, J., and Ye, J.
\newblock (2011).
\newblock A multi-task learning formulation for predicting disease progression.
\newblock In \emph{Proceedings of the 17th ACM SIGKDD International Conference
  on Knowledge Discovery and Data Mining},  814--822.

\bibitem[Zhou and Zhao(2016)]{Zhou2016-ik}
Zhou, Q. and Zhao, Q.
\newblock (2016).
\newblock Flexible clustered multi-task learning by learning representative
  tasks.
\newblock \emph{IEEE Transactions on Pattern Analysis and Machine
  Intelligence}, {\bfseries 38}\penalty0 (2), \penalty0 266--278.

\bibitem[Zou(2006)]{Zou2006-yg}
Zou, H.
\newblock (2006).
\newblock The adaptive lasso and its oracle properties.
\newblock \emph{Journal of the American Statistical Association}, {\bfseries
  101}\penalty0 (476), \penalty0 1418--1429.

\end{thebibliography}



\end{document}